\documentclass[copyright,creativecommons]{eptcs}
\providecommand{\event}{UNIF 2010} 
\usepackage{breakurl}              
\usepackage{latexsym,epsfig}
\usepackage{latexsym}
\usepackage{color}
\usepackage{graphicx}
\usepackage{verbatim}
\usepackage{amsmath,amsfonts,amssymb,mathrsfs,stmaryrd}
\usepackage[arrow,matrix,curve]{xy}
\usepackage{subfigure}
\usepackage{paralist}
\newcommand{\TechRepOnly}[1]{\color{blue}{\bf Only included in the technical report:}\\\color{blue}{#1}\color{black}}
\newcommand{\PaperOnly}[1]{\color{green}{\bf Only included in the paper:}\\\color{green}{#1}\color{black}}
\newenvironment{techreponly}{\color{blue}{\noindent\bf Only included in the technical report:}\\}{\color{black}}

\newcommand{\makePaper}{
\renewcommand{\TechRepOnly}[1]{}
\renewcommand{\PaperOnly}[1]{{##1}}

}

\makePaper

\DeclareFontFamily{OT1}{pzc}{}
\DeclareFontShape{OT1}{pzc}{m}{it}
             {<-> s * [1.12] pzcmi7t}{}
\DeclareMathAlphabet{\mathscr}{OT1}{pzc}%
                                 {m}{it}
\DeclareMathAlphabet{\mathpzc}{OT1}{pzc}{m}{it}
\renewcommand{\cal}[1]{\mathpzc{#1}}
\renewcommand{\mathcal}[1]{\mathpzc{#1}}

\newtheorem{theorem}{Theorem}[section]
\newtheorem{definition}[theorem]{Definition}
\newtheorem{example}[theorem]{Example}
\newtheorem{lemma}[theorem]{Lemma}
\newenvironment{proof}{{\it Proof.}}{\hfill$\Box$}
\newenvironment{proof*}{\proof}{}

\newcommand{\terminates}{{\hspace{-1pt}\Downarrow \hspace{-1pt}}}
\newcommand{\DIV}{{\Uparrow}}
\newcommand{\syxor}{\mathrel{|}}

\newcommand{\itbind}{\mathit{bind}}
\newcommand{\itenv}{\mathit{env}}
\newcommand{\itapp}{\mathit{app}}
\newcommand{\itlam}{\mathit{lam}}
\newcommand{\itvar}{\mathit{var}}
\newcommand{\itlet}{\mathit{let}}
\newcommand{\itFail}{\mathit{Fail}}
\newcommand{\itBind}{\mathit{Bind}}
\newcommand{\itExp}{\mathit{Exp}}
\newcommand{\itBV}{\mathit{BV}}
\newcommand{\itEnv}{\mathit{Env}}
\newcommand{\itemptyEnv}{\mathit{emptyEnv}}

\newcommand{\tletrec}{{\tt letrec}}

\newcommand{\tletrx}[2]{(\tletrec~#1 ~{\tt in}~#2)}

\newcommand{\exchainI}[2]{\{y_{i+1}=A_{i+1}[y_{i}]\}_{i=#1}^{#2}}

\newcommand{\FV}{{\mathit{FV}}}

\newcommand{\Var}{{\mathit{Var}}}

\newcommand{\chole}{[\cdot]}

\newcommand{\LNEED}{L_{\mathit{need}}}

\newcommand{\set}[1]{\{ #1 \}}

\newcommand{\iEnv}{{\mathit{Env}}}
\newcommand{\tin}{{\tt in}}

\newcommand{\BChain}{\texttt{BCh}}
\newcommand{\gentzen}[2]{{\displaystyle
    \frac{#1}{#2}}}

\newcommand{\vdiagramm}[4]{
$\xymatrix@C1cm{
\cdot \ar@{->}[d]_{#1} \ar@{->}[r]^{#2} & \cdot \ar@{-->}[d]^{#4} \\
\cdot \ar@{-->}[r]_{#3}                & \cdot
}$}
\newcommand{\ddiagramm}[4]{
$\xymatrix@C1cm@R0.5cm{
\cdot \ar@{->}[d]_{#1} \ar@{->}[r]^{#2} & \cdot \ar@{-->}[ddl]^{#4} \\
\cdot \ar@{-->}[d]_{#3}                &  \\
\cdot & \\
}$}

\newenvironment{regeln}
 {\begin{list}{}{\parskip=0.2cm \listparindent=0cm \labelsep=0.15cm \topsep=0.2cm \itemsep=0.2cm \parsep=0.1cm}}
 {\end{list}}
\newcommand{\regel}[3]{\item[\bf #1] #2 \\[0.1cm] #3}
\newcommand{\regelo}[2]{\item[\bf #1] #2}
\newcommand{\regelp}[3]{\item[\bf #1] $\phantom{.}$ \\ #2 \\[0.1cm] #3}
\newcommand{\regelq}[5]{\item[\bf #1] #2 
  \begin{minipage}{0.48\textwidth}
    \hspace{1.5cm}\textbf{#4}\hspace{0.15cm}#5
  \end{minipage}
  \\[0.1cm] #3}

\newcommand{\reg}[2]{\textbf{#1}~~#2}

\newenvironment{compactmath}
 {\begin{center}$}
 {$\end{center}}

\author{Conrad Rau\thanks{This author is supported by the DFG under grant SCHM 986/9-1.}
  ~and Manfred Schmidt-Schau{\ss}
  \institute{Institut f{\"u}r Informatik \\
    Goethe-Universit{\"a}t\\
    Postfach 11 19 32\\
    D-60054 Frankfurt, Germany\\
    \email{\{rau,schauss\}@ki.informatik.uni-frankfurt.de} } 
} 

\title{Towards Correctness of Program Transformations Through Unification and
  Critical Pair Computation}
\def\authorrunning{{\it C.~Rau and M.~Schmidt-Schau{\ss}}}
\def\titlerunning{{\it Computing Overlaps by Unification}}
 
\begin{document}
\maketitle

\begin{abstract}
Correctness of program transformations in extended lambda calculi with a
contextual semantics is usually based on reasoning about the operational
semantics which is a rewrite semantics. A successful approach to proving
correctness is the combination of a context lemma with the computation of
overlaps between program transformations and the reduction rules, and then
of so-called complete sets of diagrams.  The method is similar to the
computation of critical pairs for the completion of term rewriting systems.
We explore cases where the computation of these overlaps can be done in
a first order way by variants of critical pair computation that use
unification algorithms.  As a case study 
we apply the method to a lambda calculus with recursive let-expressions 
and describe an effective unification algorithm 
to determine all overlaps of a set of transformations with all
reduction rules.  The unification algorithm employs many-sorted terms, the
equational theory of left-commutativity modelling multi-sets, context
variables of different kinds and a mechanism for compactly representing
binding chains in recursive let-expressions.
\end{abstract}

\section{Introduction and Motivation}
 \label{sec:intro-motiv}
 
Programming languages are often described by their syntax and their
operational semantics, which in principle enables the implementation
of an interpreter and a compiler in order to put the language into
use. Of course, also optimizations and transformations into low-level
constructs are part of the implementation. The justification of
correctness is in many cases either omitted, informal or by intuitive
reasoning.  Inherent obstacles are that programming languages are
usually complex, use operational features that are not deterministic
like parallel execution, concurrent threads, and effects like input
and output, and may even be modified or extended in later releases.

Here we want to pursue the approach using contextual semantics for
justifying the correctness of optimizations and compilation and to
look for methods for automating the correctness proofs of
transformations and optimizations.

We assume given the syntax of programs ${\cal P}$, a deterministic
reduction relation $\to \;\subseteq {\cal P} \times {\cal P}$ that
represents a single execution step on programs

and values that represent the successful end of program execution.
The reduction of a program may be non-terminating due to language
constructs that allow iteration or recursive definitions.  For a
program $P \in {\cal P}$ we write $P\terminates$ if there is a
sequence of reductions to a value, and say $P$ \emph{converges} (or
terminates successfully) in this case.  Then equivalence of programs
can be defined by $P_1 \sim P_2 \iff \big(\mbox{for all } C:
C[P_1]\terminates \iff C[P_2]\terminates\big)$, where $C$ is a
context, i.e. a program with a hole $[\cdot]$ at a single position.
Justifying the correctness of a program transformation $P \leadsto
P'$ means to provide a proof that \mbox{$P \sim P'$}.  Unfortunately,
the quantification is over an infinite set: the set of all contexts,
and the criterion is termination, which is undecidable in general.
Well-known tools to ease the proofs are context lemmas
\cite{milner:77}, ciu-lemmas \cite{felleisen-hieb:92} and
bisimulation, see e.g. \cite{howe:89}.

The reduction relation $\rightarrow$ is often given as a set of rules 
$l_i \to r_i$ similarly to rewriting rules, 
but extended with different kinds of meta-variables and some other constructs,
together with a strategy determining when to use which rule and at which
position.  In order to prove correctness of a program transformation that is
also given in a rule form $s_1 \to s_2$, we have to show that $\sigma(s_1)
\sim \sigma(s_2)$ for all possible rule instantiations $\sigma$   
i.e.  $C[\sigma(s_1)]\terminates \iff C[\sigma(s_2)]\terminates$ for all
contexts $C$. Using the details of the reduction steps and induction on
the length of reductions, the hard part is to look for conflicts between
instantiations of $s_1$ and some $l_i$, i.e. to compute all the overlaps of
$l_i$ and $s_1$, and the possible completions under reduction and
transformation. This method is reminiscent of the critical pair criterion of
Knuth-Bendix method \cite{Knuth-Bendix:70} but has to be adapted to an
asymmetric situation, to extended instantiations and to higher-order terms.

In this paper we develop a unification method to compute all overlaps of left
hand sides of a set of transformations rules and the reduction rules of the
calculus $\LNEED$ which is a call-by-need lambda calculus with a
letrec-construct (see \cite{schmidt-schauss-copy-rta:07}). We show that a
custom-tailored unification algorithm can be developed that is decidable and
produces a complete and finite set of unifiers for the required equations.
The following expressiveness is required: {\em Many-sorted terms} in order to
avoid most of the junk solutions; {\em context variables} which model the
context meta-variables in the rule descriptions; {\em context classes} allow
the unification algorithm to treat different kinds of context meta-variables
in the rules; the {\em equational theory of multi-sets} models the
letrec-environment of bindings; {\em Empty sorts} are used to approximate
scoping rules of higher-order terms, where, however, only the renaming can be
modeled. Since the reduction rules are linear in the meta-variables, we
finally only have to check whether the solutions produce expressions that
satisfy the distinct variable convention.  {\em Binding Chains} in
letrec-expressions are a syntactic extension that models binding sequences 
of unknown length in the rules.  This also permits to finitely represent
infinitely many unifiers, and thus is indispensable for effectively computing
all solutions.

The required complete sets of diagrams can be computed from the
overlaps by applying directed transformations and reduction rules.
These can be used to prove correctness of program transformations by
inductive methods.

Since our case study is done for a small calculus, the demand for extending
the method to other calculi like the extended lambda calculus in
\cite{schmidt-schauss-schuetz-sabel:08} would justify further research.  

In Section~\ref{sec-calc} we present the syntax and operational
semantics of a small call-by-need 
lambda calculus with a cyclic let. The normal order reduction rules
and 
transformations are defined.  In Section~\ref{sec-encoding}, the
translation into extended first-order terms is explained.
Section~\ref{sec-unif} contains a description of the unification
algorithm that computes overlaps of left hand sides of rules and
transformations 
in a finite representation.  Finally, in Section~\ref{sec-example}, we
illustrate a run of the unification algorithm by an example.
%

\section{A Small Extended Lambda Calculus with letrec}
\label{sec-calc}

In this section we introduce the syntax and semantics of a small
call-by-need lambda calculus and use it as a case-study. Based on the
definition of the small-step reduction semantics of the calculus we
define our central semantic notion of \emph{contextual equivalence} of
calculi expressions and correctness of program transformations.  We
illustrate a method
to prove the correctness of program transformations which uses a
\emph{context lemma} and \emph{complete sets of reduction diagrams}.

\subsection{The Call-by-Need Calculus $\LNEED$}
\label{sec:call-need-calculus}

We define a simple call-by-need lambda calculus $\LNEED$ which is
exactly the call-by-need calculus of
\cite{schmidt-schauss-copy-rta:07}. Calculi that are related are in
\cite{schmidt-schauss-sabel-machkasova-rta:10},
and  \cite{ariola:97}.
  
The set ${\cal E}$ of $\LNEED$-expressions is as follows where $x, x_i$ are variables:
\begin{eqnarray*}
s_i,s,t \in {\cal E} &::=& x \syxor(s ~t) \syxor
         (\lambda x. s)~| ~\tletrx{x_1 = s_1, \ldots , x_n = s_n}{t} 
\end{eqnarray*}
We assign the names {\em application}, {\em abstraction}, or {\em letrec-expression} 
to the expressions $(s~t)$, $(\lambda x. s)$,
$\tletrx{x_1 = s_1, \ldots, x_n = s_n}{t}$, respectively. A group of
letrec-bindings, also called {\em environment}, is abbreviated as $Env$.
 
We assume that variables $x_i$ in letrec-bindings are
all distinct, that letrec-expressions are identified up to reordering of
binding-components (i.e.~the binding-components can be interchanged), and that,
for convenience, there is at least one binding.   
Letrec-bindings are recursive, i.e.,~the scope of $x_j$ in 
$\tletrx{x_1 = s_1,\ldots, x_{n-1} = s_{n-1}}{s_{n}}$ 
are all expressions $s_i$ with $1 \leq i \leq n$.
Free and bound variables in expressions and $\alpha$-renamings are defined as usual.
The set of free variables in $t$ is denoted as $\FV(t)$. We use the 
distinct variable convention (DVC), i.e., 
all bound variables in expressions are assumed to
be distinct, and free variables are distinct from bound variables.  The
reduction rules are assumed to implicitly $\alpha$-rename bound variables in
the result if necessary.

A {\em context} $C$ is an expression from $\LNEED$ extended by a symbol
$[\cdot]$, the {\em hole}, such that $[\cdot]$ occurs exactly once (as
sub-expression) in $C$.
A formal definition is:
\begin{definition}\label{def-C-context}
{\em Contexts} ${\cal C}$ are defined by the following grammar:
\begin{eqnarray*}
 C\in{\cal C} &::=& \chole \syxor(C ~s) \syxor(s ~C) 
 \syxor (\lambda x. C)
 \syxor ~\tletrx{x_1 = s_1,\ldots, x_{n} = s_{n}}{C}  
 \syxor~\tletrx{\iEnv, x=C}{s}
\end{eqnarray*}
\end{definition}
Given a term $t$ and a context $C$, we write $C[t]$ for the
$\LNEED$-expression constructed from $C$ by plugging $t$ into the
hole, i.e, by replacing $[\cdot]$ in $C$ by $t$, where this
replacement is meant syntactically, i.e., a variable capture is
permitted. Note that $\alpha$-renaming of contexts is
restricted. 


\begin{definition}
\label{def-red-rules} 
The  {\em unrestricted reduction rules} for the calculus 
$\LNEED$ are defined in Figure~\ref{figure-reductions-LNEED}. 
Several reduction rules are denoted by their name prefix, e.g.\ the union of
(llet-in) and (llet-e) is called (llet), the union of (cp-e) and (cp-in) is
called (cp), the union of (llet) and (lapp) is called (lll).
\end{definition}
\begin{figure*}[hbt]
\[\begin{array}{|ll|}\hline
\mbox{(lbeta)}&  ((\lambda x. s)~r)  \to  \tletrx{x = r}{s}\\
\mbox{(cp-in)} & \tletrx{x  = s, \iEnv~ }{C[x]}  \to \tletrx{x = s, \iEnv~ }{C[s]} \\
 & \mbox{where } s \mbox{ is an abstraction or a variable}\\
\mbox{(cp-e)} & \tletrx{x =s,\iEnv, y = C[x]}{r}~\to \tletrx{x=s,\iEnv, y = C[s]}{r} \\
 & \mbox{where } s \mbox{ is an abstraction or a variable}\\
\mbox{(llet-in)}& \tletrx{\iEnv_1}{\tletrx{\iEnv_2}{r}} \to \tletrx{\iEnv_1,\iEnv_2}{r}\\
\mbox{(llet-e)}& \tletrx{\iEnv_1, x = \tletrx{\iEnv_2}{s_x}}{r}  
     \to  \tletrx{\iEnv_1, \iEnv_2, x = s_x}{r}\\
\mbox{(lapp)} &  (\tletrx{\iEnv}{t}~s)   \to    \tletrx{\iEnv}{(t~s)} \\  
\hline
\end{array}\]
\caption{Unrestricted reduction rules of $\LNEED$ (also used as transformations)} 
\label{figure-reductions-LNEED} 
\end{figure*}

The reduction rules of $\LNEED$ contain different kinds of meta-variables. 
The meta-variables $r,s,s_x,t$ denote arbitrary $\LNEED$-expressions. 
$\iEnv,\iEnv_1,\iEnv_2$ represent letrec-environments and $x,y$ denote 
bound variables. All meta-variables can be instantiated by an
$\LNEED$-expression of the appropriate syntactical form. 
A reduction rule $\rho = l \rightarrow r$ 
is applicable to an expression $e$ if $l$ can be matched to $e$. 
Note that an expression may contain several sub-expressions that can be
reduced according to the reduction rules of Figure~\ref{figure-reductions-LNEED}. 

A standardizing order of reduction is the 
\emph{normal order reduction} (see definitions below) where reduction takes 
place only inside \emph{reduction contexts}.

\begin{definition}\label{def-R-context}
\emph{Reduction contexts} ${\cal R}$, \emph{application contexts} $\mathcal{A}$
and \emph{surface contexts} $\mathcal{S}$ are defined by the following grammars:
\[\begin{array}{lcl}
  A \in \mathcal{A}  &:= &[\cdot] \syxor  (A~s) \hspace*{1cm} \mbox{where $s$ is an expression.} \\
  R\in{\cal R} &:= & A \syxor \tletrec~\iEnv~\tin~A 
  \syxor \tletrec~y_1 = A_1, \iEnv~\tin~A[y_1] \\
  &&\syxor \tletrec~y_1 = A_1, \exchainI{1}{n}, \iEnv~\tin~A[y_n] \\
 S \in \mathcal{S}  &:= & \chole \syxor(S ~s) \syxor(s ~S) 
 \syxor ~\tletrx{y_1 = s_1,\ldots, y_{n} = s_{n}}{S}  
 \syxor~\tletrx{\iEnv, y=S}{s}
\end{array}\]
\end{definition}


A sequence of bindings of the form
$y_{m+1}=A_{m+1}[y_{m}],y_{m+2}=A_{m+2}[y_{m+1}], \ldots, y_n=A_n[y_{n-1}]$
where the $y_i$ are distinct variables, the $A_{i}$ are not the empty context
and $m < n$ is called a \emph{binding chain} and abbreviated by
$\exchainI{m}{n}$.

\begin{definition}
\label{def-no-reduction}
Normal order reduction $\xrightarrow{no}$ (called no-reduction for short)
is defined by the reduction rules in Figure~\ref{figure-no-reductions-LNEED}.  
\end{definition}
\begin{figure*}[htb]
\[\begin{array}{|ll|} \hline
  \mbox{(lbeta)}&R[(\lambda x.s)~r] \to R[\tletrec~x=r~\tin~s]\\
  \mbox{(cp-in)} & \tletrec~y=s,\iEnv~\tin~A[y] \to \tletrec~y=s,\iEnv~\tin~A[s] \\
  & \mbox{where } s \mbox{ is an abstraction or a variable.}\\
  \mbox{(cp-e)} & \tletrec~y_1 =s,y_2=A_2[y_1],\iEnv~\tin~A[y_2] \to \tletrec~y_1 =s,y_2=A_2[s], \iEnv~\tin~A[y_2]\\
\mbox{(cp-e-c)} & \tletrec~y_1 =s,y_2=A_2[y_1], \exchainI{2}{n}, \iEnv~\tin~A[y_n]\\
  &\quad\to \tletrec~y_1 =s,y_2=A_2[s], \exchainI{2}{n}, \iEnv~\tin~A[y_n]\\
  & \mbox{in the cp-e rules $s$ is an abstraction or a variable and $A_2$ is
          a non-empty context.} \\
  \mbox{(llet-in)}& \tletrx{\iEnv_1}{\tletrx{\iEnv_2}{r}}  \to  \tletrx{\iEnv_1,\iEnv_2}{r}\\
  \mbox{(llet-e)}& \tletrec~y_1 =(\tletrec~\iEnv_1~\tin~r), \iEnv_2~\tin~A[y_1]
  \to \tletrec~y_1 = r, \iEnv_1,\iEnv_2~\tin~A[y_1]\\  
  \mbox{(llet-e-c)}& \tletrec~y_1 =(\tletrec~\iEnv_1~\tin~r), \exchainI{1}{n}, \iEnv_2~\tin~A[y_n]\\
  &\quad\to \tletrec~y_1 = r, \iEnv_1, \exchainI{1}{n}, \iEnv_2~\tin~A[y_n]\\
  \mbox{(lapp)} & R[((\tletrec~\iEnv~\tin~r)~t)] \to R[(\tletrec~\iEnv~\tin~(r~t))]
  \\ \hline
\end{array}\]
\caption{Normal order reduction rules of $\LNEED$} \label{figure-no-reductions-LNEED}
\end{figure*}

Note that the normal order reduction is unique.  
A {\em weak head normal form in $\LNEED$ (WHNF)} is defined as either
an abstraction $\lambda x.s$, or an expression $\tletrx{\iEnv}{\lambda x.s}$.

The {\em transitive  closure} of the reduction relation $\to$ is denoted as
$\xrightarrow{+}$ and the {\em transitive and reflexive closure} of $\to$ is
denoted as $\xrightarrow{*}$.  
Respectively we use $\xrightarrow{no,+}$
for the transitive  closure of the normal order reduction relation, 
$\xrightarrow{no,*}$ for its reflexive-transitive closure, and 
$\xrightarrow{no,k}$ to indicate $k$ normal order reduction steps.
If for an expression $t$ there exists a (finite) sequence of normal
order reductions $t \xrightarrow{no,*} t'$ to a WHNF $t'$, we say that 
the reduction \emph{converges} and denote this as 
$t~\terminates~t'$ or as $t\terminates$ if $t'$ is not important.
Otherwise the reduction is called \emph{divergent} and we write $t\DIV$. 

The semantic foundation of our calculus $\LNEED$ is the equality of expressions
defined by contextual equivalence.  
\begin{definition}[Contextual Preorder and
  Equivalence]\label{def-contextual-equivalence}   
Let $s,t$ be $\LNEED$-expressions.  Then:
\[\begin{array}{lcll}
s \le_c t  & \text{iff} & ~\forall C:~C[s]\terminates \Rightarrow  C[t]\terminates\\
s \sim_c t &  \text{iff} & ~s \le_c t \wedge t \le_c s
\end{array} \]
\end{definition}

%

\begin{definition}
A \emph{program transformation} $T \subseteq \LNEED \times \LNEED$ is a binary
relation on $\LNEED$-expressions. 
A program transformation is called \emph{correct} iff
$T \subseteq\ \sim_c$.
\end{definition}

Program transformations are usually given in a format similarly to
reduction rules (as in Figure~\ref{figure-reductions-LNEED} and
Figure~\ref{figure-no-reductions-LNEED}). 
A program transformation $T$ is written as $s \xrightarrow{T} t$
where $s,t$ are meta-expressions i.e. expression that contain
meta-variables. 
Here we restrict our attention for the sake of simplicity to the
program transformations that are given by the reduction rules in 
Figure~\ref{figure-reductions-LNEED}.

An important tool to prove contextual equivalence is a \emph{context lemma}
(see for example \cite{milner:77},
\cite{schmidt-schauss-sabel:09},\cite{schmidt-schauss-schuetz-sabel:08}),
which allows to restrict the class of contexts that have to be considered in
the definition of the contextual equivalence from general $\cal{C}$ to
$\cal{R}$ contexts.  

However, often $\cal{S}$-contexts are more appropriate for computing
overlaps and closing the diagrams, so we will use $\cal{S}$-contexts
instead of $\cal{R}$-contexts.

\begin{lemma}
\label{lemma:contex-lemma}
Let $s,t$ be $\LNEED$-expressions and  
$S$ a context of class ${\cal S}$. 
$(S[s]\terminates \Rightarrow S[t]\terminates)$~iff~
$\forall C: (C[s]\terminates \Rightarrow C[t]\terminates)$; 
i.e. $s \leq_c t$.
\end{lemma}
\begin{proof}
A proof of this lemma when the  contexts are  
in class ${\cal R}$ is in 
\cite{schmidt-schauss-sabel:09}.  Since every ${\cal R}$-context is
also an ${\cal S}$-context, the lemma holds.
\end{proof}

\begin{figure}[ht]
\subfigure[Forking in the proof of $s \leq_c t$]{
\begin{minipage}[b]{0.23\textwidth}
$
\xymatrix@R=0.6cm@C=0.6cm{
S[s] \ar@{->}[d]_{no,k} \ar@{->}[r]^{T} & S[t] \\ 
s' & \\
}$
\label{forking-a}
\end{minipage}
}
\subfigure[Splitting the no-sequence]{
\begin{minipage}[b]{0.23\textwidth}
$
\xymatrix@R=0.6cm@C=0.6cm{
S[s] \ar@{->}[d]_{no} \ar@{->}[r]^{T} & S[t] \\ 
s_0 \ar@{->}[d]_{no,k-1} &  \\
s' & \\
}$
\label{forking-b}
\end{minipage}
}
\subfigure[Application of a forking diagram]{
\begin{minipage}[b]{0.23\textwidth}
$
\xymatrix@R=0.6cm@C=0.6cm{
S[s] \ar@{->}[d]_{no} \ar@{->}[r]^{T} & S[t] \ar@{-->}[d]_{no,*} \\
s_0 \ar@{->}[d]_{no,k-1} \ar@{-->}[r]_{T',*} & t' \\
s' & \\
}$
\label{forking-c}
\end{minipage}
}
\subfigure[Inductive proof of $s \leq_c t$]{
\begin{minipage}[b]{0.23\textwidth}
$
\xymatrix@R=0.6cm@C=0.6cm{
S[s] \ar@{->}[d]_{no} \ar@{->}[r]^{T} & S[t] \ar@{-->}[d]_{no,*} \\
s_0 \ar@{->}[d]_{no,k-1} \ar@{-->}[r]_{T',*} & t' \ar@{-->}[d]_{no,*} \\
s' \ar@{-->}[r]_{T',*} & \txt{$t''$ \\ {\tiny WHNF}} \\
}
$
\label{forking-d}
\end{minipage}
}
\caption{Sketch of the correctness proof for $s \xrightarrow{T} t$
  \label{forking-abb}}
\end{figure}
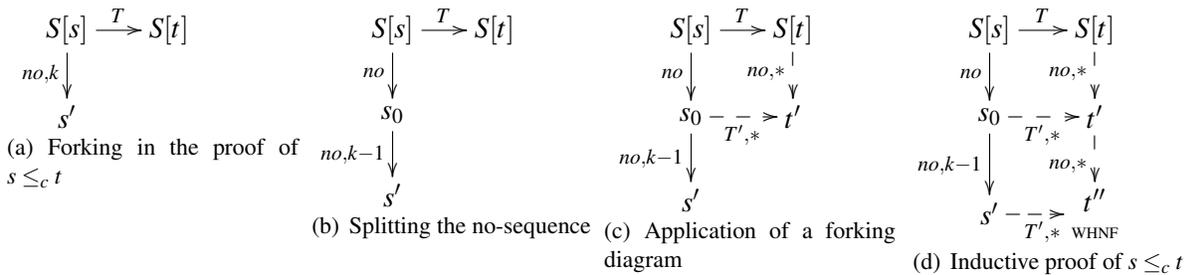

To prove the correctness of a transformation $s \xrightarrow{T} t$
we have to prove that 
$s \sim_c t \Leftrightarrow   s \leq_c t \; \wedge \; t \leq_c s$
which by Definition \ref{def-contextual-equivalence} amounts to showing 
$\forall C: C[s]\terminates \Rightarrow C[t]\terminates \; \wedge \;
C[t]\terminates \Rightarrow C[s]\terminates$. 
The context lemma yields that it is sufficient to show 
$\forall S: S[s]\terminates \Rightarrow S[t]\terminates \; \wedge \;
S[t]\terminates \Rightarrow S[s]\terminates$.
We restrict our attention here to
$S[s]\terminates \Rightarrow S[t]\terminates$ because
$S[t]\terminates \Rightarrow S[s]\terminates$ could be treated in
a similar way. 
To prove $s \sim_c t$ we assume that $s \xrightarrow{T} t$ and
$S[s]\terminates$ holds, i.e.~there is a WHNF $s'$, such that $S[s]
\xrightarrow{no,k} s'$ (see Figure~\ref{forking-a}).  It remains to show that
there also exists a sequence of normal order reductions from $S[t]$ to a WHNF.
This can often
be done by induction on the length $k$ of the given normal order reduction
$S[s] \xrightarrow{no,k} s'$ using \emph{complete sets of reduction diagrams}.
Therefore we split $S[s] \xrightarrow{no,k} s'$ into
$S[s] \xrightarrow{no} s_o \xrightarrow{no,k-1} s'$  
(see Figure~\ref{forking-b}).
Then an applicable \emph{forking diagram} defines how
the fork $s_0 \xleftarrow{no} S[s] \xrightarrow{T} S[t]$ can be closed
specifying two sequences of transformations such that a common
expression $t'$ is eventually reached: one starting from $S[t]$
consisting only of no-reductions and one starting from $s_0$ 
consisting of some other reductions 
(that are not normal order)
denoted by $T'$ in Figure~\ref{forking-c}.

A set of forking diagrams for a transformation $T$ is \emph{complete} if the
set comprises an applicable diagram for every forking situation.  If we have a
complete set of forking diagrams we often can inductively construct a
terminating reduction sequence for $S[t]$ if $S[s]\terminates$ (as indicated
in Figure~\ref{forking-d}).  To prove $S[t]\terminates \Rightarrow
S[s]\terminates$ another complete set of diagrams called \emph{commuting
  diagrams} is required which usually can be deduced from a set of forking
diagrams (see \cite{schmidt-schauss-schuetz-sabel:08}). 
We restrict our attention to complete sets of forking diagrams. 

\begin{example}
\label{example:forking-dia}
Example forking diagrams are
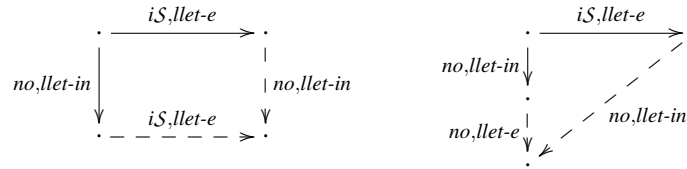
\begin{compactmath} 
\xymatrix@R=1cm{
\cdot \ar@{->}[rr]^{i\mathcal{S},\textit{llet-e}} \ar@{->}[d]_{no,\textit{llet-in}} & & 
  \cdot \ar@{-->}[d]^{no,\textit{llet-in}} \\
\cdot \ar@{-->}[rr]^{i\mathcal{S},\textit{llet-e}} & & \cdot
}
\hspace*{1cm}
\xymatrix@R=0.5cm{
\cdot \ar@{->}[rr]^{i\mathcal{S},\textit{llet-e}} \ar@{->}[d]_{no,\textit{llet-in}} & & 
  \cdot \ar@{-->}[ddll]^{no,\textit{llet-in}} \\
\cdot \ar@{-->}[d]_{no,\textit{llet-e}}                  & & \\
\cdot & & 
}
\end{compactmath}
where the dashed lines indicate existentially quantified reductions and the
prefix $i\mathcal{S}$ marks that the transformation is not a normal order
reduction (but a so called \emph{internal reduction}
which we also call transformation), 
and occurs within a surface context.  By application of the
diagram a fork between a \emph{(no,llet-e)} and the transformation
\emph{(llet-in)} can be closed.  The forking diagrams specify two reduction
sequences such that a common expression is eventually reached. 
The following reduction sequence illustrates an application of the above diagram: 
\begin{compactmath}
\begin{array}[t]{rl} 
 \multicolumn{2}{l}{
   \tletrx{\iEnv_{1}, x=\tletrx{\iEnv_{2} }{s} }{\tletrx{ \iEnv_{3} }{r}} }\\ \hline 
 \xrightarrow{no,\textit{llet-in}} & \tletrx{\iEnv_{1},\iEnv_{3}, x=\tletrx{\iEnv_{2} }{s}}{r} \\ 
 \xrightarrow{i{\cal S} \vee no,\textit{llet-e}} & \tletrx{\iEnv_{1},\iEnv_{3},\iEnv_{2}, x=s }{r} \\ 
 \multicolumn{2}{l}{\quad\mbox{the last reduction is either an no-reduction if $r = A[x]$, otherwise it is an internal reduction}} \\ \hline
 \xrightarrow{i{\cal S},\textit{llet-e}}&\tletrx{ \iEnv_{1},\iEnv_{2}, x=s }{\tletrx{ \iEnv_{3} }{r}} \\ 
 \xrightarrow{no,\textit{llet-in}}& \tletrx{ \iEnv_{1},\iEnv_{2},\iEnv_{3}, x=s }{r} \\
\end{array}
\end{compactmath}

The square diagram covers the case, where \emph{(no,llet-in)} is followed by
an internal reduction. The triangle diagram covers the other case, where the
reduction following \emph{(no,llet-in)} is \emph{(no,llet-e)}.  One can view
the forking diagram as a description of local confluence.

\end{example}

The computation of a complete set of diagrams by hand is cumbersome and
error-prone. Nevertheless the diagram sets are essential for proving
correctness of a large set of program transformations in this setting.  For
this reason we are interested in automatic computation of complete diagram
sets.  


The first step in the computation of a complete set of forking diagrams for a
transformation $\mathit{T}$ is the determination of all forks of the
form $\xleftarrow{no,\mathit{red}} \cdot \xrightarrow{i{\cal S},T}$ where
$\mathit{red}$ is an no-reduction and $T$ is not a normal order reduction 
(but a transformation in an ${\cal S}$-context).  
Such forks are given by \emph{overlaps} between no-reductions
and the transformation.  Informally we say that 
$\mathit{red}$ and $T$ overlap in an expression $s$ if $s$ contains a normal
order redex $\mathit{red}$ and a $T$ redex (in a surface context). To find an
overlap between an no-reduction $\mathit{red}$ and a transformation 
$\mathit{T}$ it is sufficient, by definition 
of the normal order reduction, to determine all surface-positions in
$\mathit{red}$ where a $\mathit{T}$-redex can occur. For the computation of
all forks we have to consider only \emph{critical overlaps} where an overlap
does not occur at a variable position (Example~\ref{example:forking-dia}
illustrates such a critical overlap).  Forks stemming from non-critical
overlaps at variable positions can always be closed by a predefined set of standard diagrams. All
(critical) overlaps between no-reductions and a given transformation 
$T$ can be computed by a variant of critical pair computation based on unification.
The employed unification procedure will be explained in the next section.

\section{Encoding Expressions as Terms in a Combination of Sorted
  Equational Theories and Context}\label{sec-encoding}
In this section we develop a unification method to compute proper overlaps
for forking diagrams.
According to the context lemma for
surface contexts (Lemma \ref{lemma:contex-lemma}) we restrict the overlaps to
the transformations applied in surface contexts.  A complete description of a
single overlap is the unification equation $S[l_{T,i}] \doteq l_{no,j}$, where
$l_{T,i}$ is a left hand side in Figure \ref{figure-reductions-LNEED}, and
$l_{no,j}$ a left hand side in Figure \ref{figure-no-reductions-LNEED}, and
$S$ means a surface context.
To solve these unification problems we translate 
the meta-expressions from transformations and no-reduction rules into
many sorted terms with some special constructs to mirror the syntax
of the reduction rules in the lambda calculus. 
The constructs are \begin{inparaenum}[i)]
\item context variables of different context classes ${\cal A},{\cal S}$ and ${\cal C}$,
\item a left-commutative function symbol $\itenv$ to model that
  bindings in letrec-environments can be rearranged
\item a special construct $\BChain(\ldots)$ to represent binding chains
  of variable length as they occur in no-reduction rules.
\end{inparaenum}

The presented unification algorithm is applicable to terms with
the mentioned extra constructs.
We do not use the general unification combination algorithms in
\cite{msscomb:89,baaderschulz:92-short}, since we only have a special theory
\emph{LC} that models multi-sets of bindings in letrec-environments of our
calculus, and moreover, it is not clear how to adapt the general combination
method to context classes and binding chains.
 
\subsection{Many Sorted Signatures, Terms and Contexts}
Let $\mathcal{S} = \mathcal{S}_1 \uplus \mathcal{S}_2$ be the disjoint union
of a set of theory-sorts $\mathcal{S}_1$ and a set of free sorts
$\mathcal{S}_2$. We assume that $\itExp$ is a sort in $\mathcal{S}_2$.
Let $\Sigma = \Sigma_1 \uplus \Sigma_2$ be a many-sorted signature of (theory-
and free) function symbols, where every function symbol comes with a fixed
arity and with a single sort-arity of the form $f: S_1 \times \ldots \times
S_n \to S_{n+1}$, where $S_i$ for $i = 1,\ldots,n$ are the argument-sorts and
$S_{n+1} $ is called resulting sort. For every $f \in \Sigma_i$ for $i = 1,2$
the resulting sort must be in $\mathcal{S}_i$.  Note, however, that there may
be function symbols $f \in \Sigma_i$ that have argument-sorts from
$\mathcal{S}_j$, for $i \not= j$.
There is a set $\mathcal{V}^0$ of first-order variables that are 0-ary and
have a fixed sort and are ranged over by $x,y,z,\ldots$, perhaps with indices.
We write $x^S$ if the variable $x$ has the sort $S$.
There is also a set $\mathcal{V}^1$ of context-variables which are unary and
are ranged over by $X,Y,Z$, perhaps with indices.  We assume that for every
sort $S$, there is an infinite number of variables of this sort, and that
there is an infinite number of context variables of sort $\itExp \to
\itExp$.  Let $\mathcal{V} = \mathcal{V}^0 \cup \mathcal{V}^1$.  The set
of terms $\mathcal{T}(\mathcal{S},\Sigma,\mathcal{V})$ is the set of terms
built according to the grammar
$x~|~f(t_1,\ldots,t_n)~|~X(t)$, where sort conditions are obeyed.  Let
$\Var(t)$ be the set of first-order variables that occur in $t$ and let
$\Var^1(t)$ be the set of context variables that occur in $t$.  A context $C$
is a term in 
$\mathcal{T}(\itExp,\Sigma \cup [\cdot], \mathcal{V})$
such that there is exactly one occurrence of a the special hole constant
$[\cdot]$ in the context and the sort at the position of the
hole is $\itExp$.
 
A term $s$ without occurrences of variables is called {\em ground}. We also
allow sorts without any ground term, also called {\em empty sorts}, since this
is required in our encoding of bound variables. The term $s$ is called {\em
 almost ground}, if for every variable $x$ in $s$, there is no function
symbol in $\Sigma$ where the resulting sort is the sort of $x$, and hence no ground term of this sort.

A substitution $\sigma$ is a mapping $\sigma: \mathcal{V} \to
\mathcal{T}(\mathcal{S},\Sigma,\mathcal{V}^0)$, such that $\sigma(x^S)$ is a
term of sort $S$ and $\sigma(X)$ is a context. As usual we extend $\sigma$ to
terms, where every variable $x$ in a term is replaced by $\sigma(x)$.

\subsection{Encoding of $\LNEED$-Expressions as Terms}
\label{subsubsec:encoding}
The sort and term structure according to the expression structure of the
lambda calculus $\LNEED$ (from section \ref{sec:call-need-calculus}) is as
follows.  There are the following sorts: $\itBind, \itEnv, \itExp, \itBV$,
for bindings, environments, expressions and bound variables, respectively; where
$\mathcal{S}_1 = \{\itEnv\}$ and $\mathcal{S}_2 = \{\itBind, \itExp,\itBV\}$.
There are the following  function symbols: 

\begin{compactmath}
\begin{array}{l|ll}
\mbox{theory function symbols ($\Sigma_1$)} & \mbox{free function symbols ($\Sigma_2$)} \\\hline
\begin{array}{l}
  \itemptyEnv::\phantom{\itBind\times\itEnv\to}\itEnv\\
   \itenv::\phantom{empty}\itBind \times \itEnv \to \itEnv\\
 \end{array}
&
  \begin{array}{l}
 \itlet::~~\itEnv \times    \itExp \to \itExp\\
 \itapp::\itExp \times \itExp \to \itExp\\ 
 \itlam::\itBV \times  \itExp \to \itExp
 \end{array}
&
 \begin{array}{l}
  \itbind::\itBV \times \itExp \to \itBind\\
  \itvar::\itBV \phantom{.\times \itExp} \to \itExp \\
 \end{array}
 \end{array}
\end{compactmath}

Note that there are free function symbols that map from $\itEnv$ to $\itExp$,
but there is no free function symbol that maps to $\itEnv$.  Note also that
there is no function symbol with resulting sort $\itBV$, hence this is an
empty sort, and every term of sort $\itBV$ is a variable.

It is convenient to have a notation for nested $\itenv$-expressions:
$\itenv^*(\{t_1, \ldots, t_m\}\cup r)$ denotes the term
$\itenv(t_1,\itenv(t_2, \ldots, \itenv(t_m,r)\ldots))$, where $r$ is not of
the form $\itenv(s,t)$. Due to our assumptions on terms of sort $\itEnv$ and
the sort of context variables, only the constant $\itemptyEnv$ and variables
are possible for $r$.

As an example  the expression $\tletrx{x = \lambda y. y, z = x~x}{z}$  is encoded as\\
$\itlet(\itenv^*(\{\itbind(x,\itlam(y,\itvar(y))),\itbind(z,\itapp(\itvar(x),\itvar(z)))\}\cup \itemptyEnv),\itvar(z))$,
where $x,y,z$ are variables of sort $\itBV$. 
 
To model the multi-set property of letrec-environments, i.e., that bindings can be
reordered, we use the equational theory {\em left-commutativity} ($LC$) with
the following axiom:
$\itenv(x,\itenv(y,z)) = \itenv(y,\itenv(x,z))$ (for the $LC$-theory and
unification modulo LC see
\cite{dovier-pontelli:06,
  dantsin-voronkov:99}).  The
equational theory $LC$ is a congruence relation on the terms, which is denoted
as $=_{\mathit{LC}}$.
The {\em pure equational theory} is defined as restricted to the axiom-signature, i.e. to the terms 
$\mathcal{T}(\{\itEnv,\itBind\},\Sigma_1, \mathcal{V}_{\itEnv} \cup \mathcal{V}_{\itBind})$, 
where $\mathcal{V}_S$ is the set of variables of sort $S$. 
The {\em combined equational theory} is defined on the set of terms
$\mathcal{T}(\mathcal{S},\Sigma, \mathcal{V}^0)$.  Note that it is a disjoint
combination w.r.t. the function symbols, but not w.r.t. the sorts.

The following facts about the theory $LC$ can easily be verified:
\begin{lemma} For the equation theory $LC$, the following holds in
  $\mathcal{T}(\mathcal{S},\Sigma,\mathcal{V}^0)$:
\label{lemma:prob-eq-theory}
\begin{compactitem}
\item The terms in the $LC$-axioms are built only from
  $\Sigma_1$-symbols and variables, and the axioms relate two terms of
  equal sort which must be in $\mathcal{S}_1$.
 
\item For every equation $s =_{LC} t$, the equality $\Var(s)=\Var(t)$ holds.
\item The equational theory ${LC}$ is non-collapsing, i.e, there is no
  equation of the form $x =_{LC} t$, where $t$ is not the variable $x$.
\item If $C[s] =_{LC} t$ and $s$ has a free function symbol as top
  symbol, then there is a context $C'$ and a term $s'$ such that 
  $C[s]=_{LC} C'[s'], C' =_{LC} C$, $s =_{LC} s'$ and $C'[s'] = t$.  
  This follows from general properties of combination of equational
  theories and properties of the theory LC.
\item The equational theory ${LC}$ has a finitary and decidable
  unification problem 
  (see\cite{dovier-pontelli:06,
    dantsin-voronkov:99}).
\end{compactitem}
\end{lemma}

In order to capture binding chains of variable length as they occur 
in the definition of the no-reduction rules (Figure~\ref{figure-no-reductions-LNEED})
the syntax construct $\BChain(N_1,N_2)$ is introduced, where $N_i$ are  integer 
variables that can be instantiated
with $N_1 \mapsto n_1$, $N_2 \mapsto n_2$, where \mbox{$0 < n_1 < n_2$}.
An instance $\BChain(n_1,n_2)$ for $n_1,n_2 \ge 1$ represents the 
following binding chain:
  $\itbind(y_{n_1+1},A_{n_1+1}(\itvar(y_{n_1})))$, $\itbind(y_{n_1+2},A_{n_1+2}(\itvar(y_{n_1+1}))),\ldots,  
   \itbind(y_{n_2},A_{n_2}(\itvar(y_{n_2-1})))$,
where  the names $y_i,A_i$   are reserved for these
purposes and are all distinct. The $\BChain$-expressions are permitted only in the
$\itenv^*$-notation, like a sub-multi-set, and we denote this for example as
$\itenv^*(\ldots \cup\BChain(N_1,N_2) \cup r)$.

Context-classes are required to correctly model the overlappings in
$\LNEED$.  The transformations in Figure~\ref{figure-reductions-LNEED}
contain only $C$-contexts, whereas in
Figure~\ref{figure-no-reductions-LNEED} there are also $\mathcal{A}$-
and $\mathcal{R}$-contexts, and the overlapping also requires surface
contexts $\mathcal{S}$.  The grammar definition of $\cal{A}$-,
$\cal{R}$- and $\cal{S}$-contexts (definition~\ref{def-R-context})
justifies the replacement of $\mathcal{R}$-contexts by expressions
containing only $\mathcal{A}$-contexts and
$\BChain$-expressions. Thereby some rules of Figure
\ref{figure-no-reductions-LNEED} may be split into several rules.  The
context class $\mathcal{C}$ means all contexts and
$\mathcal{S}$ 
means all contexts where the hole is not in an abstraction.  In the
term encoding, these translate to context
variables.  
The unification algorithm must know how to deal with context variables
of classes $\mathcal{A}$, $\mathcal{S}$ and $\mathcal{C}$.
The partial order on context classes is $\mathcal{A} < \mathcal{S} < \mathcal{C}$. 
For every almost ground context $C$ it can be decided whether $C$
belongs to $\mathcal{A}$ (or $\mathcal{S}$, respectively).  We will
use the facts that equational deduction w.r.t. $LC$ does not change
the context class of almost ground contexts, and that prefix and
suffix contexts of almost ground contexts $C$ have the same context
class as $C$ (among $\mathcal{A}$, $\mathcal{S}$ and $\mathcal{C}$).

 \section{A Unification Algorithm LCSX for Left-Commutativity, Sorts and Context-Variables}
\label{sec-unif}

We define unification problems and solutions as extension of
equational unification (see \cite{baadersnyder:2001}).

A {\em unification problem} is a pair $(\Gamma,\Delta)$, where $\Gamma
= \{s_1 \doteq t_1,\ldots, s_n \doteq t_n\}$, the terms $s_i$ and
$t_i$ are of the same sort for every $i$ and may also contain
$\BChain$-expressions, every context variable is labelled with a
context class symbol, and $\Delta = (\Delta_1,\Delta_2)$ is a
constraint consisting of a set of context variables $\Delta_1$ and a
set $\Delta_2$ of equations and inequations of the form $N_i+1 = N_j$
and $N_i < N_j$ for the integer variables $N_i$.  The intention is
that $\Delta_1$ consists of context variables that must not be
instantiated by the empty context, and that the constraints $\Delta_2$
hold for $\sigma(N_i)$ after instantiating with $\sigma$.

A \emph{solution} $\sigma$ of $(\Gamma,\Delta)$, with 
$\Gamma = \{s_1 \doteq t_1,\ldots, s_n \doteq t_n\}$  
is a substitution $\sigma$ according to the following conditions:
\begin{inparaenum}[i)] 
\item it instantiates variables by terms, context variables by
  contexts of the correct context class that are nontrivial if
  contained in $\Delta_1$, and the integer variables $N_i$ by positive
  integers according to the constraint $\Delta_2$.
\item $\sigma(s_i), \sigma(t_i)$ are almost ground for all $i$. It is
  assumed that the $\BChain$-constructs $\BChain(n_1,n_2)$ are
  expanded into a binding chain as explained above,
\item $\sigma(s_i) =_{LC} \sigma(t_i)$ for all $i$.
\end{inparaenum}

A unification problem $\Gamma$ is called {\em almost linear}, if every context
variable occurs at most once and every variable of a non-empty sort occurs at
most once in the equations.
 
\begin{definition}
\label{def:init-fork-prob}
Let $\Pi_T$ be the set of left hand sides of reduction rules from
Figure~\ref{figure-reductions-LNEED} and $\Pi_{no}$ the set of left hand sides of
no-reduction rules from Figure~\ref{figure-no-reductions-LNEED} where the
reduction contexts $R$ in (lbeta) and (lapp) are instantiated by the four
possibilities for $R$:
\begin{inparaitem}[]
\item $A$,
\item $\tletrx{\iEnv}{A}$,
\item $\tletrx{y_1 = A,\iEnv}{A_2[y_1]}$,
\item $\tletrx{y_{N_1} = A,\BChain(N_1,N_2),\iEnv}{A[y_{N_2}]}$ with constraint $N_1 < N_2$.
\end{inparaitem}
The meta-variable $s$ in the cp rules 
(that can be either a variable or an abstraction) is instantiated by
\begin{inparaitem}[]
\item $\itvar(z)$ and 
\item an abstraction $\lambda x.t$ where $t$ denotes a meta-variable for an
  arbitrary expression.
\end{inparaitem}
With $\Pi_T',\Pi_{no}'$ we denote the sets where left hand sides  
of rules are encoded as terms. 

We consider the set of unification problems $\Gamma_i =
\set{S(l_{T,i}) \doteq l_{no,j} ~|~ l_{no,j} \in \Pi_{no}' }$ with
$l_{T,i} \in \Pi_T'$ and $S$ is a surface context variable.  The sets
$\Pi_T'$ and $\Pi_{no}'$ are assumed to be variable disjoint, which
can be achieved by renaming.  The initial set $\Delta_1$ of context
variables only contains the $A_2$-context from the
\mbox{(cp-e)}-reductions, and $\Delta_2$ may contain some initial
constraints from the rules.  The pairs $(\Gamma_{i},\Delta)$ are
called the {\em initial $\LNEED$-forking-problems}.
\end{definition}

Note that initial $\LNEED$-forking-problems are almost linear, there is at
most one $\BChain$-construct, which is in the environment of the topmost
let-expression, and there are no variables of type $\itBind$.

\begin{definition}
A {\em final unification problem} $S$ of an initial $\Gamma$ is a set of
equations $s_1 \doteq t_1,\ldots, s_n \doteq t_n$, such that 
$S = S_{\itBV} \cup S_{\neg\itBV}$, and every equation in $S_{\itBV}$ is of
the form $x \doteq y$ where $x,y$ are of sort $\itBV$ and every equation in
$S_{\neg\itBV}$ is of the form $x \doteq t$, where $x$ is not of sort $\itBV$,
and the equations in $S_{\neg\itBV}$ are in DAG-solved form.
\end{definition}

Given a final unification problem $S$, the represented solutions $\sigma$ could be
derived by first instantiating the integer variables, expanding the
$\BChain$-constructs into binding chains, instantiating all context variables
and variables that are not of sort $BV$ and then turning the equations 
into substitutions.  

A final unification problem $S$ derived from $\Gamma$ {\em satisfies the
  distinct variable convention} (DVC), if for every derived solution $\sigma$,
all terms in $\sigma(\Gamma)$ satisfy the DVC. This property is decidable: 
If $t_1 \doteq t_2$ is the initial problem, then apply  the substitution
$\sigma$ derived from $S$ to $t_1$.
The DVC is violated if the following condition holds: Let $M_{\itBV}$
be the set of $\itBV$-variables occurring
in 
$\sigma(t_1)$.  For every $\BChain$-construct $\BChain(N_1,N_2)$
occurring in 
$\sigma(t_1)$ we add the variable $y_{N_2}$ to $M_{\itBV}$.
If 
$\sigma(t_1)$ makes two variables in $M_{\itBV}$ equal, then the DVC
is violated, and the corresponding final problem is discarded.

\begin{example}  Unifying (the first-order encodings of) 
$\lambda x. \lambda y. x$ and $\lambda u. \lambda v. v$, 
the unification succeeds and generates an instance that represents
$\lambda x. \lambda x. x$, which does not satisfy the DVC.  
Thus a variant of our unification can  efficiently check
alpha-equivalence of lambda-expressions that satisfy the DVC. 
\end{example}

We proceed by describing a unification algorithm starting with initial
$\LNEED$-unification problems $(\Gamma,\Delta)$. It is intended to be
complete for all common instances that represent $\LNEED$-expressions that
satisfy the DVC, i.e. where all bound variables are distinct and the bound
variables are distinct from free variables. Final unification problems
that lead to expressions that do not satisfy the DVC are discarded. 

Given an initial unification problem $\Gamma = \{s_1 \doteq
t_1\};\Delta$, the (non-deterministic) unification algorithm described
below will non-deterministically compute a final unification problem
$S$ 
or fail. A finite complete set of final unification problems can be
attained by gathering all final unification problems in the whole tree
of all non-deterministic choices.
We implicitly use symmetry of $\doteq$ if not stated otherwise.
We divide $\Gamma$ in a solved part $S$, (a final unification problem), 
and a still to be solved part $P$.
We usually omit $\Delta$ in the notation if it is not changed by the rule.
\\[0.2cm]\noindent
{\bf Standard unification rules.}  \\[0.3cm]
\reg{Dec}
{$\gentzen{S;\;\{f(s_1,\ldots,s_n) \doteq  f(t_1,\ldots,t_n)\} \uplus P} 
   {S;\; \{s_1 \doteq   t_1, \ldots, s_n \doteq   t_n \} \cup P}$}
 \quad If $f$ is a free function symbol (i.e. $f \neq \itenv$). \\[0.2cm]
\begin{minipage}{0.4\textwidth}
\reg{Solve}
{$\gentzen{S;\; \{x \doteq  t\} \uplus P}{\{x \doteq t\} \cup S;\; P}$}
\end{minipage} 
\begin{minipage}{0.4\textwidth}
\reg{Trivial}{$\gentzen{S; \; \{s \doteq  s\} \uplus P}{S;\; P}$}
\end{minipage} \\[0.2cm]
\begin{minipage}{0.4\textwidth}
\reg{Fail}
{$\gentzen{S;\; \{f(\ldots) \doteq  g(\ldots)\} \uplus P}{\itFail}$}
\end{minipage} 
\begin{minipage}{0.6\textwidth}
\reg{DVC-Fail}{$\gentzen{~~S;\;\emptyset~~}{\itFail}$}
{\quad
  \begin{minipage}{0.65\textwidth}
  If $S$ is final and the DVC is violated w.r.t. the initial problem.
\end{minipage}
}
\end{minipage}
\\[0.2cm]
\indent Note that the occurs-check is not necessary, since $P$ is
almost linear and an equation $x \doteq t$ for variables $x$ of type
$\itBV$ implies that $t$ is a variable.
\\[0.2cm]\noindent
{\bf Solving equations with context variables.}
The rules for terms with contexts as top symbol using their context
classes are as follows:
The following rule operates on  context variables at any position: 
\begin{regeln}
\regelo{Empty-C \label{empty-c}}
 {$\gentzen{S;\; P;\;  \Delta_1 ~~~  }
     {\begin{array}{l}
        \mbox{select one of the following possibilities} \\
         S;\; P;\; \{X\} \cup \Delta_1  \quad \mbox{or} \quad
         \{X \mapsto [\cdot]\} \cup S;\; \{X \mapsto [\cdot]\}P;\;\Delta_1
       \end{array}}$}
 \quad If $X$ occurs  in $P$ and $X \not\in \Delta_1$.
\end{regeln}
Assume there is an equation $X(s) \doteq t$, where the top symbol of
$t$ is not a context variable and $X \in \Delta_1$. Note that the sort
of $X(s)$ is $\itExp$.  There are the following possibilities:
\begin{regeln}
\regel{Dec-CA \label{dec-ca}}
 {$\gentzen{S;\; \{X(s) \doteq \itapp(t_1,t_2)\} \uplus P}
   {\{X \mapsto \itapp(X',t_2)\} \cup S;\; \{X'(s) \doteq t_1\} \cup P}$}
 {$X'$ is a fresh context variable of the same context class as $X$.}
\regel{Dec-CC \label{dec-cc}}
 {$\gentzen{S;\; \{X(s) \doteq f(t_1,t_2)\} \uplus P}{\{X \mapsto f(t_1,X')\} \cup S; \; \{X'(s) \doteq t_2\} \cup P}$
 \hspace*{1cm} if  $t_2$ is of sort $\itExp$.}
 {$X'$ is a fresh context variable of the same context class as $X$
  (it may only be $\mathcal{C}$ or $\mathcal{S}$) and 
  $f$ is a function symbol such that
  $f \in \{\itlet, \itapp\}$.}
\regel{Dec-CL \label{dec-cl}} 
 {$\gentzen{S;\; \{X(s) \doteq \itlet(t_1,t_2)\} \uplus P}
   {\{X \mapsto \itlet(\itenv^*(\{\itbind(x,X')\} \cup z),t_2)\} \cup S;\; 
     \{\itenv^*(\{\itbind(x,X'(s))\} \cup z) \doteq t_1\} \cup P}$}
 {If $X$ is of context class $\mathcal{S}$ or $\mathcal{C}$.
  $X'$ is a fresh context variable of the same context class as $X$.}
\regel{Dec-Lam \label{dec-lam}}
 {$\gentzen{S;\; \{X(s) \doteq \itlam(t_1,t_2)\} \uplus P}{\{X \mapsto \itlam(t_1,X')\} \cup S; \; \{X'(s) \doteq t_2\} \cup P}$}
 {If $X$ is of class $\mathcal{C}$. 
  $X'$ is a fresh context variable of the class $\mathcal{C}$.}
\regelq{Fail-Lam} 
 {$\gentzen{S;\; \{X(s) \doteq \itlam(t_1,t_2)\} \uplus P}{\mathit{Fail}}$}
 {If $X$ is of class $\mathcal{A}$ or $\mathcal{S}$.}
 {Fail-Var}
 {{$\gentzen{S;\; \{X(s) \doteq \itvar(x)\} \uplus P}{\itFail}$}}
\end{regeln} 
Given an equation $X(s) \doteq Y(t)$, with $X,Y \in \Delta_1$, let
${\cal D}$ be the smaller one of the context classes of $X,Y$. Then
select one of the following possibilities:
\begin{regeln}
\regel{Merge-P \label{merge-p}}
 {$\gentzen{S;\; \{X(s) \doteq Y(t)\} \uplus P}{\{Y \mapsto ZY', X \mapsto Z\} \cup S;\; \{s \doteq Y'(t)\} \cup P}$}
 {$Y'$ is a fresh context variable of the same context class as $Y$,
  and $Z$ has context class ${\cal D}$.}
\regel{Merge-FA \label{merge-fa}}
 {$\gentzen{S;\; \{X(s) \doteq Y(t)\} \uplus P}
   {\{X \mapsto Z(\itapp(X', Y'(t)), Y \mapsto Z(\itapp(X'(s),Y'))\} \cup S;\; P}$}
 {If exactly one of the context classes of $X,Y$ is  $\mathcal{A}$.
  W.l.o.g. let  $X$ be of context class $\mathcal{A}$.
  $X', Y'$ are fresh context variables of the same context class as
  $X,Y$, respectively, and $Z$ is a fresh context variable of
  context class $\mathcal{A}$.}
\regelp{Merge-FC \label{merge-fc}}
 {\hspace*{-1cm} $\gentzen{S;\; \{X(s) \doteq Y(t)\} \uplus P}
  {\begin{array}{l}
      \mbox{choose either of the following possibilities} \\
    \{X \mapsto Z(\itapp(X',Y'(t))), Y \mapsto Z(\itapp(X'(s),Y'))\} \cup S;\; P \\
    \{X \mapsto Z(\itlet(\itenv^*(\{\itbind(x,X')\}\cup ,z),Y'(t))),Y \mapsto Z(\itlet(\itenv^*(\{\itbind(x,X'(s))\}\cup z),Y'))\} \cup S;\; P \\
    \{X \mapsto Z(\itlet(\itenv^*(\{\itbind(x,X'),\itbind(y,Y'(t))\}\cup z),w)), \\
    \phantom{\{} Y \mapsto Z(\itlet(\itenv^*(\{\itbind(x,X'(s)),\itbind(y,Y')\}\cup z),w))\} \cup S;\; P 
    \end{array}}$}
 {\hspace*{-0.8cm}
  \begin{minipage}{1\textwidth}
  If the context classes of $X,Y$ are different from $\mathcal{A}$.
  $X', Y'$ are fresh context variables of the same context class 
  as $X,Y$, respectively and $Z$ is a fresh context variable of context class ${\cal D}$.
  The variables $w,x,y,z$ are also fresh and of the appropriate sort.
 \end{minipage}}
\end{regeln}

\noindent
{\bf Rules for Multi-Set Equations.} 
The following additional (non-deterministic) unification rules  are sufficient to solve nontrivial 
equations of type $\itEnv$, i.e. proper multi-set-equations,
which must be of the form $\itenv^*(L_1\cup r_1)  \doteq \itenv^*(L_2\cup r_2)$, where $r_1,r_2$ are variables or the constant $\itemptyEnv$.
We will use the notation  $L$ for sub-lists in $\itenv^*$-expressions and
the notation $L_1 \cup L_2$ for union. 
In the terms $\itenv^*(L\cup t)$, we assume that
$t$ is not of the form $\itenv(\ldots)$. It is also not of the form $X(\ldots)$ due to the sort assumptions. 
Other free function symbols are disallowed, hence $t$ can only be a variable or the constant $\itemptyEnv$.
The components in the multi-set may be expressions of type $\itBind$, i.e., variables or expressions with top symbol $\itbind$,
or a  $\BChain(\ldots)$-component that represents several terms of type $\itBind$.
We also use the  convention that in the  conclusions of the rules an empty  environment $\itenv^*(\{\ \} \cup r)$ 
without any bindings and just a variable $r$ is identified with $r$.
Note that the lists allow multi-set operations like reorderings. 

Due to the initial encoding of reduction rules, if a $\BChain(N_1,N_2)$-construct occurs in a term in $P$, it occurs in an $\itenv^*$-list, 
hence there is also 
a binding $y_{N_1} = s$ in the $\itenv^*$-list, and the list is terminated with a variable derived from the environment-variable $\iEnv$.
In equations, the $\BChain(\ldots)$-components initially appear only on one side, which cannot be changed by the unification.
Also the $\itenv^*$-list is an immediate sub-term of a top let-expression, 
which may change after applying unification rules.
Due to these conditions, 
we  assume that the left term in the equation does not contain $\BChain(\ldots)$-components. 

If there is an equation $\itenv^*(L_1 \cup r_1)  \doteq \itenv^*(L_2 \cup r_2)$, then 
select one of the following possibilities:
\begin{regeln}
\regelo{Solve-E \label{solve-e}}
 {$\gentzen{S;\; \{\itenv^*(L_1 \cup r_1) \doteq \itenv^*(L_2\cup r_2)\} \uplus P}
   {\{r_1 \mapsto \itenv^*(L_2\cup z_3), r_2 \mapsto \itenv^*( L_1\cup z_3)\} \cup S;\; P}$}
 \quad If $r_1,r_2$ are variables; $z_3$ is a fresh variable.
\regel{Dec-E \label{dec-e}}
 {$\gentzen{S;\; \{\itenv^*(L_1\cup r_1) \doteq  \itenv^*(L_2\cup r_2)\} \uplus P}
   {S;\; \{t_1 \doteq t_2,  \itenv^*(L_1\setminus\{t_1\}\cup r_1) \doteq  \itenv^*(L_2\setminus\{t_2\}\cup r_2)\} \uplus P}$}
 {If $L_1$ and $L_2$ contain binding expressions $t_1$ and, $t_2$.}
\regelp{Dec-Ch \label{dec-ch}}
 {\hspace*{-1cm}
  $\gentzen{S;\; \{\itenv^*(L_1\cup r_1) \doteq  \itenv^*(\BChain(N_1,N_2) \cup L_2\cup r_2)\} \uplus P;\; (\Delta_1,\{N_1 < N_2\}\cup \Delta_2)}
  {\begin{array}{l} 
      \mbox{select one of the following possibilities} \\
      (i)\; S;\; 
      \{t_1 \doteq \itbind(y_{N_2}, A_{N_2}(\itvar(y_{N_1}))),\\
      \phantom{S;\;\{} 
      \itenv^*(L_1 \setminus \{t_1\} \cup r_1) \doteq \itenv^*(L_2 \cup r_2)\} \cup P;\; 
      \{A_{N_2}\} \cup \Delta_1,\{N_1+1 = N_2\}\cup \Delta_2 \\
      (ii)\; S;\; 
      \{t_1 \doteq \itbind(y_{N_3}, A_{N_3}(\itvar(y_{N_1}))), \\
      \phantom{S;\;\{} 
      \itenv^*(L_1 \setminus \{t_1\} \cup r_1) \doteq  \itenv^*(\BChain(N_3,N_2) \cup  L_2 \cup r_2)\} \cup P;\; 
      \{A_{N_3}\} \cup \Delta_1,\{N_1+1 = N_3,N_3 < N_2\}\cup \Delta_2 \\
      (iii)\; S;\;
      \{t_1 \doteq \itbind(y_{N_2}, A_{N_2}(\itvar(y_{N_3}))), \\
      \phantom{S;\;\{} 
      \itenv^*(L_1 \setminus \{t_1\} \cup r_1) \doteq \itenv^*(\BChain(N_1,N_3) \cup L_2 \cup r_2)\} \cup P;\; 
      \{A_{N_2}\} \cup \Delta_1,\{N_1 < N_3,N_3+1 = N_2\}\cup \Delta_2 \\
      (iv)\; S;\; 
      \{t_1 \doteq \itbind(y_{N_4}, A_{N_4}(\itvar(y_{N_3}))), \\
      \phantom{S;\;\{} 
      \itenv^*(L_1 \setminus \{t_1\} \cup r_1) \doteq \itenv^*(\BChain(N_1,N_3) \cup  \BChain(N_4,N_2)\cup L_2\cup r_2)\} \cup P;\\
      \phantom{S;\;} 
      \{A_{N_4}\} \cup \Delta_1,\{N_1 < N_3,N_3+1 = N_4,N_4 < N_2\}\cup \Delta_2 \\ 
   \end{array}}$}
 {\hspace*{-0.8cm}
   Where $y_{N_2},y_{N_3},y_{N_4},A_{N_2},A_{N_3},A_{N_4},N_3,N_4$ are fresh variables of appropriate sort.}
%
\regel{Fail-E \label{fail-e}}
 {$\gentzen{S;\; \{\itenv^*(L \cup t) \doteq \itemptyEnv\} \uplus P}{\itFail}$.}
 {If $L$ is nonempty, i.e contains at least one binding or at least one $\BChain$-expression.}
\end{regeln}

An invariant of the rules that deal with $\BChain$ is that the variables $N_i$
may appear at most twice in $\Gamma$; at most twice explicit in $\Delta_2$ and at most once
in $\BChain$-expressions.

\subsection{Properties of the LCSX-Unification Algorithm}
\begin{lemma}
For initial problems, the algorithm LCSX  terminates.
\end{lemma}
\begin{proof} 
For this we can ignore the rules that change $\Delta$. 

The following measure is used, which is a lexicographical
combination of several component measures: $\mu_1$ is the number of
occurrences of $\itlet$ in $P$; the second component $\mu_2$ is the
following size-measure, where $\itenv^*(L \cup r)$ has measure
$7m+m'+\sum\mu_2(t_i) +\mu_2(r)$ where $m$ is the number of
$\itbind$-expressions in $L$ and and $m'$ is the number of
$\BChain$-expressions in $L$.

The critical applications are the guessing rules for equations with
top-context variables, and the rules for multi-equations.  The context
variable-guessing either decreases the size or the number of
occurrences of let.  The multi-equation rules in rule {\bf Dec-Ch} 
have to be analyzed. The new constructed bind-term has size 5, so the
subcases $(i)$ -- $(iii)$ strictly reduce the size.  The subcase
$(iv)$ adds $6$ to the size due to new sub-terms, and removes $7$
since $t_1$ is a non-$\BChain$-expression and removed from the
multi-set.
\end{proof}

\begin{lemma}
The non-deterministic rule-based unification algorithm LCSX is sound
and complete in the following sense: every computed final
unification problem 
that leads to an expression satisfying the DVC represents a set of
solutions 
and every solution of the initial unification problem that
represents an expression satisfying the DVC is represented by one
final system of equations.
\end{lemma}
\begin{proof}
Soundness can be proved by standard methods, since rules are either instantiations or instantiations using the theory LC.

Completeness can be proved, if every rule is shown to be complete, and if
there are no stuck unification problems that have solutions.  
The \textbf{Solve} rules are complete since solved variables 
(in equations of the form $x \doteq t$) are just marked as such, i.e. moved 
to a set of solved equations.
Solving equations $X(s) \doteq t$ is complete: if $t$ is a variable, then it can be
replaced; if $t$ is a proper term of type $\itExp$, then all cases are covered
by the rules.  In the case that the equation is $X(s) = Y(t)$, the rules are
also complete, and also respect the context classes of $X,Y$.  If the equation
is $s \doteq s$, then it will be removed, and if it is of the form $f(\ldots)
\doteq f(\ldots)$ then decomposition applies. In the case that the top symbol
is $\itenv$, the rules for multi-equations apply, i.e., the rules for
$\itenv^*$. Using the properties of the equational theory $\mathit{LC}$ and
the considerations in \cite{dovier-pontelli:06
}), we see
that the rules are complete.
\end{proof}

\begin{theorem}
The rule-based algorithm LCSX terminates if applied to initial
$\LNEED$-forking-problems. Thus it decides unifiability of these sets of
equations. Since it is sound and complete, and the forking possibilities are
finite, the algorithm also computes a finite and complete set of final 
unification problems by gathering all possible results.
\end{theorem}

\begin{theorem}
The computation of all overlaps between the rules in Figure
\ref{figure-reductions-LNEED} and left hand sides of normal order reductions
in Figure \ref{figure-no-reductions-LNEED} can be done using the algorithm
LCSX. The unification algorithm terminates in all of these cases and
computes a finite set of final unification problems 
and hence all the critical pairs w.r.t. our normal order reduction.
\end{theorem}

\section{Running the  Unification Algorithm LCSX}\label{sec-example}
\begin{example} 
The goal is to compute a complete set of forks 
for the transformation (cp-e) 
\begin{compactmath}
  \tletrx{x=s,\iEnv, z = C[x]}{r}~\to \tletrx{x=s,\iEnv, z = C[s]}{r}
\end{compactmath}
from Figure~\ref{figure-reductions-LNEED}. We instantiate the meta-variable
$s$ by the expression $\lambda w.t$ and translate the left hand side of the
rule into the term language, resulting in the following initial forking
problem to be solved
\begin{compactmath} 
  \set{S(\itlet(\itenv^*(\{\itbind(x,\itlam(w,t)),\itbind(z ,C(\itvar(x)))\}\cup \iEnv),r))\doteq l_{no,j}}.
\end{compactmath}
where $l_{no,j}$ is an encoded left hand side of an no-reduction rule.
We pick a single equation from this set: 
\begin{compactmath}
  \begin{array}{rl}
    & S(\itlet(\itenv^*(\{\itbind(x,\itlam(w,t)), \itbind(z ,C(\itvar(x)))\}\cup \iEnv_1),r)) \\
    \doteq & \itlet(\itenv^*(\{\itbind(x',\itlam(w',t')),\itbind(y_{N_1},A_{N_1}(\itvar(x')))\}
    \cup \BChain(N_1,N_2) \cup \iEnv_2),A(y_{N_2}))
  \end{array}
\end{compactmath}
which describes the overlaps between the (cp-e) transformation and the normal
order (cp-e-c) reduction.  No we compute one possible final problem via
the presented unification algorithm. A nontrivial possibility is to choose $S =
[\cdot]$ via the \textbf{Empty-C}-rule and then using decomposition for
$\itlet$ which leads to $r = A(y_{N_2}))$ and the equation
\begin{compactmath}
\begin{array}{ll}
       & \itenv^*(\{\itbind(x,\itlam(w,t)), \itbind(z ,C(\itvar(x)))\}\cup \iEnv_1) \\
\doteq & \itenv^*(\{\itbind(x',\itlam(w',t')),\itbind(y_{N_1},A_{N_1}(\itvar(x')))\}\cup \BChain(N_1,N_2)\cup \iEnv_2).
\end{array}
\end{compactmath}
One choice for the next step (via the rule \textbf{Dec-Ch}) results in the equations:
\begin{compactmath}
\begin{array}{ll}
  &  \itbind(z,C(\itvar(x)) \doteq \itbind(y_{N_4},A_{N_4}(\itvar(y_{N_3}))), \quad
         \itenv^*(\{\itbind(x,\itlam(w,t))\}\cup \iEnv_1) \\
\doteq & \itenv^*(\{\itbind(x',\itlam(w',t')),\itbind(y_{N_1},A_{N_1}(\itvar(x')))\} \cup (\BChain(N_1,N_3) \cup \BChain(N_4,N_2) \cup\iEnv_2)
\end{array}
\end{compactmath}
where one binding is taken from the $\BChain(N_1,N_2)$-construct and the chain
is split around this binding into two remaining chains.  The two bindings
$\itbind(x,\itlam(w,t))$ and $\itbind(x',\itlam(w',t'))$ are unified (via
\textbf{Dec-E}) and then we solve the equation between the environments
(\textbf{Solve-E}) and (after three additional \textbf{Dec}-steps, two for
\textit{bind} and one for \textit{lam}) we arrive at the system
\begin{compactmath}
\begin{array}{l}
  C(\itvar(x)) \doteq A_{N_4}(\itvar(y_{N_3})), z \doteq y_{N_4}, 
  x \doteq x', w \doteq w', t \doteq t', 
  \iEnv_2 \doteq \itenv^*(\{\itbind(x,\itlam(w,t))\}\cup \iEnv_3), \\
  \iEnv_1 \doteq \itenv^*(\{\itbind(x',\itlam(w',t')),\itbind(y_{N_1},A_{N_1}(\itvar(x')))\} \cup \BChain(N_1,N_3) \cup \BChain(N_4,N_2) \cup \iEnv_3).
\end{array}
\end{compactmath}
Next we apply \textbf{Merge-FA} to the first equation, yielding
\begin{compactmath}
 C \doteq Z(\itapp(A_{N_2}'(\itvar(y_{N_3})),C')),
 A_{N_2} \doteq Z(\itapp(A_{N_2}',C'(\itvar(x))))
\end{compactmath}
where $Z,A_{N_2}'$ are of context class ${\cal A}$ and $C'$ is of 
context class ${\cal C}$.
The final representation is:
\begin{compactmath}
\begin{array}{l}
S_{\neg\itBV} = \set{
S \doteq [\cdot], 
r \doteq A(y_{N_2}), 
C \doteq Z(\ldots),
t \doteq t', 
A_{N_2} \doteq Z(\ldots),
\iEnv_2 \doteq \itenv^*(\ldots ), 
\iEnv_1 \doteq \itenv^*(\ldots )}  \\
S_{\itBV} = \set{
z \doteq y_{N_4}, 
x \doteq x', 
w \doteq w'}
\end{array} 
\end{compactmath}
 The resulting expression is:  
\begin{compactmath}
\begin{array}{l}
\tletrec~ x'=(\lambda w'.t'), y_{N_1}=A_{N_1}[x'], \exchainI{N_1}{N_3}, \\
\phantom{\tletrec~}y_{N_4}=Z[\itapp(A_{N_2}'(\itvar(y_{N_3})),C'[x'])],\exchainI{N_4}{N_2},\iEnv_2~\mathtt{in}~A[y_{N_2}]
\end{array}
\end{compactmath}
The corresponding fork is given by reducing the expression with (no,cp-e-c) and (cp-e) respectively
\[\xymatrix{
\txt{\small $\tletrec~ x'=(\lambda w'.t'), y_{N_1}=A_{N_1}[x'],$ \\ 
   \small $\exchainI{N_1}{N_3},y_{N_4}=Z[\itapp(A_{N_2}'[\itvar(y_{N_3})],C'[x'])],$ \\
   \small $\exchainI{N_4}{N_2},\iEnv_2~\mathtt{in}~A[y_{N_2}]$}
\ar@{->}[d]_{no,\textit{cp-e-c}} \ar@{->}[dr]^{i{\cal S},\textit{cp-e}} &  \\
\txt{\small $\tletrec~ x'=(\lambda w'.t'), y_{N_1}=A_{N_1}[\mathbf{\lambda w'.t'}],$ \\ 
   \small $\exchainI{N_1}{N_3},y_{N_4}=Z[\itapp(A_{N_2}'[\itvar(y_{N_3})],C'[x'])],$ \\
   \small $\exchainI{N_4}{N_2},\iEnv_2~\mathtt{in}~A[y_{N_2}]$}
&
\txt{\small $\tletrec~ x'=(\lambda w'.t'), y_{N_1}=A_{N_1}[x'],$ \\ 
   \small $\exchainI{N_1}{N_3},$ \\
    \small $y_{N_4}=Z[\itapp(A_{N_2}'[\itvar(y_{N_3})],C'[\mathbf{\lambda w'.t'}])],$ \\
   \small $\exchainI{N_4}{N_2},\iEnv_2~\mathtt{in}~A[y_{N_2}]$}
}\]
This fork can be closed by the sequence $\xrightarrow{i{\cal S},\textit{cp-e}}
\cdot \xleftarrow{no,\textit{cp-e-c}}$.  Notice that for the determination of all
forks it is sufficient to compute final systems. The (possibly infinite) set 
of ground solutions is not required.
\end{example}

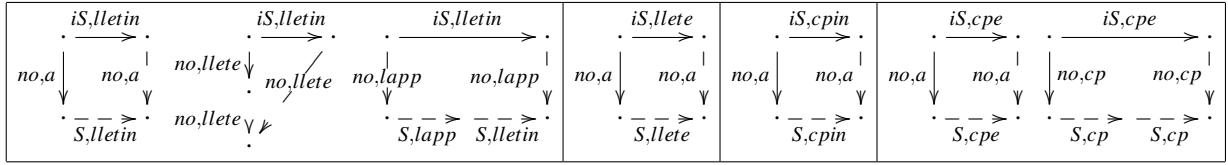
\begin{figure}[htb]
$
\setlength{\arraycolsep}{0.5mm}
\begin{array}{|ccc|c|c|cc|} \hline
\xymatrix@C0.8cm@R0.7cm{
\cdot \ar@{->}[d]_{no,a} \ar@{->}[r]^{iS,lletin} & \cdot \ar@{-->}[d]_{no,a} \\
\cdot \ar@{-->}[r]_{S,lletin}                & \cdot
} &
\xymatrix@C0.8cm@R0.35cm{
\cdot \ar@{->}[d]_{no,llete} \ar@{->}[r]^{iS,lletin} & \cdot \ar@{-->}[ddl]|(0.41){no,llete}\\
\cdot \ar@{-->}[d]_{no,llete}                &  \\
\cdot & \\
} &
\xymatrix@C0.8cm@R0.7cm{
\cdot \ar@{->}[d]|{no,lapp} \ar@{->}[rr]^{iS,lletin} & & \cdot \ar@{-->}[d]_{no,lapp} \\
\cdot \ar@{-->}[r]_{S,lapp} & \ar@{-->}[r]_{S,lletin} & \cdot \\
} &
\xymatrix@C0.8cm@R0.7cm{
\cdot \ar@{->}[d]_{no,a} \ar@{->}[r]^{iS,llete} & \cdot \ar@{-->}[d]_{no,a} \\
\cdot \ar@{-->}[r]_{S,llete}                & \cdot
} &
\xymatrix@C0.8cm@R0.7cm{
\cdot \ar@{->}[d]_{no,a} \ar@{->}[r]^{iS,cpin} & \cdot \ar@{-->}[d]_{no,a} \\
\cdot \ar@{-->}[r]_{S,cpin}                & \cdot
} &
\xymatrix@C0.8cm@R0.7cm{
\cdot \ar@{->}[d]_{no,a} \ar@{->}[r]^{iS,cpe} & \cdot \ar@{-->}[d]_{no,a} \\
\cdot \ar@{-->}[r]_{S,cpe}                & \cdot
} &
\xymatrix@C0.8cm@R0.7cm{
\cdot \ar@{->}[d]^{no,cp} \ar@{->}[rr]^{iS,cpe} & & \cdot \ar@{-->}[d]_{no,cp} \\
\cdot \ar@{-->}[r]_{S,cp} & \ar@{-->}[r]_{S,cp} & \cdot \\
} \\ 
\hline 
\end{array}
$
\caption{Complet sets of forking diagrams for llet and cp transformations.
\label{fig-complet-forking-set}}
\end{figure}

We implemented the presented unification algorithm LCSX in Haskell to
compute all forks between transformations and no-reductions.  The
program computes 1214 overlaps for
the $\LNEED$ calculus, and also searches for closing reduction sequences.
Via this
method we 
were able to close (almost\footnote{Some simple commuting diagrams
  for cp reductions are not automatically closed, due to renaming of
  bound variables.}) all forks. 
The complete sets of forking diagrams 
for the
transformations llet and cp is in
Figure~\ref{fig-complet-forking-set} 
The implementation is available at:
\url{http://www.ki.informatik.uni-frankfurt.de/research/dfg-diagram/en}.
More informaiton can be found in \cite{rau-schmidt-schauss-IB41:10}.

\section{Conclusion and Further Work}
We have provided an 
method using first-order unification with equational theories, sorts,
context variables and context classes and 
binding chains of variable length to compute all critical overlaps between a
set of transformation rules and a set of normal order rules in a
call-by-need lambda calculus with letrec-environments.  Further work
is to apply this method to further transformations and also to extend
the method in order to make it applicable to other program calculi as
in \cite{schmidt-schauss-schuetz-sabel:08}, where variable-variable
bindings are present in the 
rules, and to calculi with
data structures and case-expressions.

\end{document}